\documentclass[sigconf]{acmart}
\usepackage[utf8]{inputenc}
\usepackage{graphicx}
\usepackage{longtable}
\usepackage{hyperref}
\usepackage{caption}
\usepackage[brazil]{babel}

\hypersetup{
    colorlinks=true,
    linkcolor=blue,
    filecolor=magenta,      
    urlcolor=cyan,
    pdftitle={Overleaf Example},
    pdfpagemode=FullScreen,
    }

\title{ 
Understanding the relationships between the perceptions of burnout and instability in Software Engineering }
\begin{document}
\author{Danilo Monteiro Ribeiro}
\email{danilo.ribeiro@zup.com.br}
\affiliation{%
  \institution{Zup Innovation}
  \city{São Paulo}
  \state{São Paulo}
  \country{Brazil}
}
\affiliation{%
  \institution{Faculdade Senac}
  \city{Recife}
  \state{Pernambuco}
  \country{Brazil}
}
\begin{abstract}
Changes are common during the development process of software. These changes can increase the perception of instability of software developers. Understand the relationship between human factors and Software Engineering's process is important to try to mitigate and prevent problems. One of these factors is burnout. Burnout is considered a disease that can impact several factors in Software Engineering, such as productivity, turnover, but mainly the health of the developers. Therefore, understanding the relationship between instability and burnout can help those involved to prevent and strategize so that developers can feel better and, consequently, produce more. The main objective of this work is to identify and describe the relationship between instability and burnout in the perception of members who participate in the software development process. To reach this objective, we conducted a cross-section survey with 411 respondents, which had its sampling by convenience and self-selection. In addition to identifying relationships between variables, confirmatory factor analysis techniques were also applied. The main results of this work are: The relationship between the perception of burnout and team, technological and task instability, which is positive and significant for the dimensions of exhaustion and cynicism. For the dimension of efficacy, it was negative and weak with technological and team instabilities and does not exist with task instability. In addition, exhaustion is perceived more frequently by respondents. The instability with the highest perception of frequency by the practitioners is that of the task. The results found are important because both the market and the academy can guide their efforts to reduce the perception of burnout and instability in software engineers. Furthermore, new research can be carried out to verify the impact of instability on developers, bringing a new perspective to monitor the perception of instability in the context of software development.

\end{abstract}
\maketitle
\section*{Publication Information}

This work was originally written in Portuguese and has been translated into English using GPT to provide an English-language version while maintaining the original structure and content as closely as possible. The original article was published in the following venue:  

\textbf{SBES '22: Proceedings of the XXXVI Brazilian Symposium on Software Engineering, Pages 58-67}.  

The original paper can be accessed at the ACM Digital Library via the following link:  

\url{https://dl.acm.org/doi/10.1145/3555228.3555251}

Please cite the original article when referring to this work.

\section{Introduction}

Human aspects play a crucial role in Software Engineering, as they help us understand how software engineers respond to workplace demands and how to create a more suitable work environment \cite{LENBERG201515}. 

One of these aspects is \textit{burnout}, a syndrome in which work-related stressors negatively impact task performance and overall well-being \cite{maslach2001job}. 

The consequences of \textit{burnout} include destructive behaviors such as interpersonal conflicts, exhaustion, and undesirable organizational outcomes, including high turnover rates, absenteeism, and decreased work performance \cite{swider2010born}. 

\textit{Burnout} has been widely investigated across different disciplines, particularly in health sciences, due to its impact on work performance and the increasing number of diagnosed cases. A Gallup report, based on a sample of 7,500 workers from various sectors, indicates that approximately 23\% experience \textit{burnout} symptoms on a daily basis \cite{wigert2018employee}. In specific professions, such as healthcare (e.g., doctors and nurses), this percentage can reach up to 82\% \cite{lima2018prevalencia}. 

Workplace health can be severely affected by \textit{burnout}. According to Salvagioni et al. \cite{salvagioni2017physical}, \textit{burnout} increases health risks such as type 2 diabetes, cardiovascular diseases, depression, insomnia, substance abuse, and gastrointestinal disorders. Recognizing its severity, the World Health Organization (WHO) officially classified \textit{burnout} as an occupational disease in the ICD-11 (International Classification of Diseases).

Beyond being a public health issue, \textit{burnout} also has significant economic consequences. A study by Sicking \cite{sicking2011burnout} revealed that approximately 10\% of medical leave days in Germany were caused by \textit{burnout}, leading to high costs and reduced productivity.

In the field of computing, Huarng \cite{huarng2001burnout} investigated \textit{burnout} in the software industry and found that at least 39\% of respondents exhibited high levels in at least one of the dimensions of \textit{burnout}, suggesting a concerning prevalence. A similar trend was observed by Cook~\cite{cook2015job} in a later study conducted in 2015.

A key aspect of work-related \textit{burnout} is that it results from an individual’s perception of their work environment and how they interpret workplace events. According to Maslach and Leiter \cite{maslach2016burnout}, environmental changes contribute significantly to this perception. This is particularly relevant in Software Engineering, where change is considered an inherent part of the development process \cite{williams2003guest}. 

Changes can occur in various ways within a software team’s environment, including personnel turnover, evolving requirements, programming language updates, and schedule pressures \cite{boehm2008making, maurer2006agile}. In this study, we define the perception of such changes in terms of three dimensions: team instability, technological instability, and task instability.

Given that changes are common in software development, that the perception of environmental instability may contribute to \textit{burnout}, and that \textit{burnout} is a pressing societal issue affecting software professionals, an important question arises:  

\begin{center}
    \textbf{RQ1 - What is the relationship between perceptions of burnout and environmental instability in Software Engineering?}
\end{center}

To answer this research question, we conducted a **cross-sectional survey** with 411 participants. The results indicate weak to moderate correlations between \textit{burnout} dimensions and perceived instability, except for task instability and the efficacy dimension of \textit{burnout}, which showed no significant relationship.

Additionally, we obtained descriptive insights into \textit{burnout} and instability in Software Engineering. For instance, exhaustion emerged as the most frequently experienced negative dimension among respondents. Meanwhile, efficacy—measured inversely, meaning higher levels indicate lower \textit{burnout}—was generally high among software professionals.

Regarding environmental instability, task instability was perceived as the most common challenge in software development, followed closely by team instability and technological instability, with nearly identical levels of occurrence.

The remainder of this paper is structured as follows: Section~\ref{background} presents the theoretical background, including key concepts and related work. Section~\ref{chap:hypothesis} introduces the research hypotheses. Section~\ref{chap:method} details the study design and methodology. Section~\ref{results} presents the findings, including both descriptive and relational results. Section~\ref{discussion} discusses the study’s implications and provides recommendations. Finally, Section~\ref{conclusion} offers concluding remarks, discusses limitations, and highlights potential threats to validity.

\section{Theoretical Framework}
\label{background}

\subsection{Burnout in the Workplace}

According to Maslach, Schaufeli, and Leiter \cite{maslach2001job}, \textit{burnout} is a syndrome that emerges as a chronic response to work-related stressors, particularly interpersonal stressors in professional settings. It occurs when individuals struggle to adapt to workplace demands, whether due to misalignment between personal and organizational expectations or difficulties in adjusting to job-related challenges \cite{maslach2016burnout}. Consequently, affected individuals may develop \textit{burnout} syndrome.

Maslach \cite{maslach1976burned} conceptualizes \textit{burnout} as comprising three interrelated but independent dimensions:

\textbf{Emotional exhaustion} is the first dimension, characterized by a lack of enthusiasm, reduced energy, and a sense of depletion. Individuals experiencing this condition feel frustrated and tense, believing they can no longer invest the necessary effort into their work.

\textbf{Depersonalization}, also referred to as cynicism, is the second dimension. It occurs when individuals start treating colleagues, clients, and other stakeholders impersonally, displaying emotional detachment. This often arises as a response to emotional exhaustion, leading to disengagement from interpersonal relationships in the workplace \cite{maslach2016burnout}.

\textbf{Reduced personal accomplishment}, also known as low professional efficacy, is the third dimension. It reflects a tendency for individuals to self-evaluate negatively, experiencing dissatisfaction with their professional growth and a sense of incompetence in performing their job responsibilities \cite{maslach2016burnout}. As a result, affected individuals develop a persistent belief that they lack the necessary skills and abilities to execute their work effectively.

\subsection{Perception of Instability}

This study considers three categories of instability proposed by Kude et al. \cite{kude2014adaptation}: task instability, team instability, and technological instability.

\textbf{Task instability} results from the introduction of new requirements, reprioritization of tasks, and \textit{ex-post} modifications. In this study, it is defined as the perception of individuals regarding changes in their tasks, particularly in terms of evolving requirements, shifting priorities, and deadline adjustments \cite{kude2014adaptation}.

\textbf{Team instability} refers to fluctuations in team composition. According to Kude et al. \cite{kude2014adaptation}, a stable team consists of members who have worked together for an extended period. Slotegraaf and Atuahene-Gima \cite{slotegraaf2011product} define team stability as the continuity of key members throughout a project, from its initiation to product release. In this study, team instability is defined as an individual's perception of changes in team composition, specifically in terms of member departures and arrivals that impact team activities.

\textbf{Technological instability}, also referred to as technological disruption, is the third category examined. According to Kude et al. \cite{kude2014adaptation}, technological instability arises from the introduction of new technological elements, such as programming languages, frameworks, and APIs, or from external disruptions, such as modifications to third-party software that affect development processes. Additionally, it may stem from technological turbulence, which includes challenges related to development environments and platforms.

\subsection{Related Work}

Several studies in Software Engineering have explored \textit{burnout} and its implications.

Sonnetag and Brodbeck \cite{sonnentag1994stressor} found a negative correlation between the time spent learning on the job and depersonalization, as well as a negative correlation between communication levels and depersonalization. Additionally, they identified several relationships between environmental factors (e.g., job complexity and work control) and \textit{burnout} in Software Engineering.

Moore \cite{moore2000one} investigated the impact of emotional exhaustion on turnover intention in Software Engineering. The study found a positive relationship between these two variables. Furthermore, Moore observed that autonomy and rewards were negatively correlated with emotional exhaustion, while work overload and role conflict were positively associated with emotional exhaustion.

Singh and Suar \cite{singh2013health} examined the consequences of \textit{burnout} on software engineers’ health, identifying a positive correlation between \textit{burnout} and conditions such as anxiety, depression, and social dysfunction. Their findings underscore the negative impact of \textit{burnout} and the importance of addressing it.

Cook \cite{cook2015job} analyzed the prevalence of \textit{burnout} dimensions in Software Engineering. The study found that cynicism was the most prevalent dimension (43\%), followed by emotional exhaustion (32

Mellblom and Arason \cite{mellblom2019connection} explored the relationship between personality traits and \textit{burnout}, finding that individuals with higher levels of neuroticism were more prone to \textit{burnout}. However, the study did not specify which \textit{burnout} dimensions were linked to personality traits.

While this research shares similarities with previous studies on \textit{burnout} in Software Engineering, it extends beyond them by investigating the relationship between \textit{burnout} and a characteristic inherent to software projects: change.

\section{Research Hypotheses}

Based on the research question formulated, it is necessary to establish hypotheses to guide the study appropriately.

According to Maslach \cite{maslach2016burnout}, \textit{burnout} can be influenced by environmental changes. In Software Engineering, change is a frequent and inevitable phenomenon \cite{williams2003guest,boehm2008making,maurer2006agile}. Thus, it is reasonable to assume that individuals who perceive their environment as constantly changing may also perceive themselves as experiencing higher levels of \textit{burnout}.

Consequently, the primary hypothesis of this study is:

\begin{center}
    \textbf{H1 - There is a relationship between the perception of instability and burnout.}
\end{center}

Since both variables investigated (\textit{burnout} and instability) can be divided into distinct dimensions, the primary hypothesis has been subdivided to provide a more detailed understanding of these relationships. Thus, this research examines the following sub-hypotheses:

\begin{itemize}
    \item \textbf{H1.1} Task instability perception is positively related to exhaustion (1.1a) and cynicism (1.1b) and negatively related to efficacy (1.1c) in Software Engineering.
\end{itemize}

\begin{itemize}
    \item \textbf{H1.2} Team instability perception is positively related to exhaustion (1.2a) and cynicism (1.2b) and negatively related to efficacy (1.2c) in Software Engineering.
\end{itemize}

\begin{itemize}
    \item \textbf{H1.3} Technological instability perception is positively related to exhaustion (1.3a) and cynicism (1.3b) and negatively related to efficacy (1.3c) in Software Engineering.
\end{itemize}

\section{Methodological Procedures}
\label{chap:method}

\subsection{Research Approach}

This study adopts a positivist stance. According to Easterbrook et al. \cite{easterbrook2008selecting}, positivism asserts that knowledge should be based on logical inference derived from a set of observable facts. Furthermore, Easterbrook et al. \cite{easterbrook2008selecting} state that positivist researchers seek to understand reality through relationships between variables, employing statistical methods to identify such relationships and make inferences about a phenomenon based on a population sample.

Therefore, this research follows a **quantitative approach** and employs a **survey study** with a relational design to address the research question:

\begin{center}
    \textit{\textbf{What is the relationship between perceptions of burnout and project instability in Software Engineering?}}
\end{center}

\subsection{Research Design}

This study follows a **cross-sectional** (\textit{cross-section}) design, as data were collected at a single point in time to provide an overview of the current state of the field \cite{shull2007guide}.

This design was selected because it is widely used in other research fields, such as psychology, and is considered appropriate for describing relationships between variables \cite{olsen2004cross, louttit1947use, litwin2003assess, irwing2018wiley}.

\subsection{Participants and Sample}

The target population of this study consists of Portuguese-speaking Software Engineers actively engaged in software development teams. A Software Engineer, for the purposes of this study, may include developers, database analysts, software managers or project leaders, analysts, test engineers, and other professionals involved in software development.

The questionnaire was self-administered, meaning that participants completed it independently. The sampling method employed was **self-selection**, where individuals voluntarily chose to participate. Additionally, the sample was **non-probabilistic and convenience-based** \cite{hair2009analise}.

A total of 486 individuals participated in the survey. However, not all responses were included in the analysis. The following exclusion criteria were applied:
\begin{itemize}
    \item Participants who identified solely as computer science professors or from unrelated fields.
    \item Participants with no prior experience in software development.
    \item Incomplete responses.
\end{itemize}

After applying these criteria, **411 valid responses** remained for statistical analysis.

Descriptive statistics of the sample are presented in Section~\ref{sec:resultados}.

\subsection{Data Collection}

Participants were recruited through social media platforms, including Twitter, Facebook, and LinkedIn. These platforms were selected for convenience, as they were the primary networks available to the researchers for sharing the study.

The recruitment strategy involved posting the survey link and tagging influential figures in the Brazilian computing community, requesting them to share the post. Data collection occurred between **October and November 2019**.

No financial incentives were offered to participants.

\subsection{Data Analysis}

Data analysis was conducted using two statistical programming languages: **Python and R**. 

Python, along with the **Pandas** library and the **Pingouin** statistical package \cite{vallat2018pingouin}, was used for descriptive and correlational analyses.

R \cite{rlanguage} was used for **Cronbach’s Alpha calculations**, **Confirmatory Factor Analysis (CFA)**, and questionnaire validation procedures.

\subsubsection{Confirmatory Factor Analysis (CFA)}

CFA is a statistical technique used to assess the fit of a theoretical model and the correlational structure of the collected data \cite{costa2011mensuraccao}. It is also employed to evaluate the validity of a measurement scale \cite{hair2009analise}.

By conducting CFA, the researcher tests a **predefined theoretical model**, which may be derived from exploratory factor analysis or prior research, to determine whether the observed data support the hypothesized structure \cite{maroco2010analise}. In other words, CFA helps ensure that the theoretical model is appropriately represented in the collected data.

In this study, CFA was performed using R with the **lavaan** package (\textit{latent variable analysis}) \cite{beaujean2014latent}, which relies on covariance-based analysis.

To assess CFA quality, three criteria were examined \cite{maroco2010analise}:
\begin{itemize}
    \item **Model fit quality (GOF)**: Evaluates whether the model aligns with the theoretical framework.
    \item **Convergent validity**: Assesses whether scale items accurately represent their intended constructs.
    \item **Discriminant validity**: Determines whether the analyzed constructs are distinct from one another.
\end{itemize}

\subsection{Measurement Instruments}

Two validated instruments were used in this study, as described below.

\subsubsection{Burnout}

To measure **burnout**, we used the translated and validated scale for Software Engineering by Da Silva et al. \cite{da2016preliminary}. This scale was originally developed by Maslach \cite{maslach1976burned} as a general measure applicable to various professions.

The instrument has been employed in previous Software Engineering research, such as the study by Santos et al. \cite{santos2016building}. It consists of **16 items**, divided into three dimensions: 
\begin{itemize}
    \item Cynicism (4 items),
    \item Emotional exhaustion (6 items),
    \item Reduced professional efficacy (6 items, measured inversely).
\end{itemize}

Responses were recorded using a **seven-point Likert scale**, ranging from **"Never" to "Every day"**. The full scale is available at the following \href{https://docs.google.com/document/d/e/2PACX-1vQR1gD_Tb2tWgJC6bxy_yHJIYjnFieO38qnGfh3YNSBF1zrnvysVVYVv017POOrEdQK9WRo2yJtV5Au/pub}{\textit{link}}.

A **Confirmatory Factor Analysis (CFA)** was conducted to assess scale validity. The results indicated good model fit:  
($ \chi^2$= 626.157; gl = 126.000; p = 0.000; $ \chi^2$/gl = 4.90; TLI = 0.903; CFI = 0.903; IFI = 0.893; RMSEA = 0.091; SRMR = 0.078).  
Although RMSEA slightly exceeded the recommended threshold (0.089), some authors suggest more lenient cutoffs, such as 0.1 \cite{kenny2015performance, maccallum1996power}. Additionally, all factor loadings exceeded 0.5, supporting scale validity.

Cronbach’s Alpha reliability tests were also conducted:  
\begin{itemize}
    \item Emotional exhaustion: $\alpha$ = 0.91
    \item Cynicism: $\alpha$ = 0.87
    \item Professional efficacy: $\alpha$ = 0.86
\end{itemize}

These values indicate strong internal consistency.

\subsubsection{Instability}

The **Instability Scale** used in this study was developed by Ribeiro~\cite{ribeiro2020relaccoes}, based on the categorization by Kude et al. \cite{kude2014adaptation}. It measures:
\begin{itemize}
    \item Task instability,
    \item Technological disruption,
    \item Team instability.
\end{itemize}

The instrument consists of **seven-point Likert scale** items (1 = "Never"; 7 = "Very frequently"). The complete scale is available at the following \href{https://docs.google.com/document/d/e/2PACX-1vQR1gD_Tb2tWgJC6bxy_yHJIYjnFieO38qnGfh3YNSBF1zrnvysVVYVv017POOrEdQK9WRo2yJtV5Au/pub}{\textit{link}}.

CFA results supported the validity of this scale:  
($ \chi^2$= 87.249; gl = 32.000; p = 0.000; $\chi^2$/gl = 2.702; TLI = 0.951; CFI = 0.979; IFI = 0.972; RMSEA = 0.064; SRMR = 0.042).

\subsection{Ethical Considerations}

This study complied with Brazilian research ethics regulations (Resolution 466/12 - CNS-MS). Informed consent was obtained from all participants, and data collection was anonymous and confidential.

\section{Results}
\label{sec:results}

\subsection{Descriptive Results}
\label{sec:descriptive_results}

This section presents the descriptive results of the sample and the two main variables analyzed: \textit{burnout} and instability.

\subsubsection{Sample Characteristics}

Table~\ref{tab:profpop} presents the descriptive statistics of the sample regarding professional experience (years), role experience (years), months in the current project, and age. Respondents were asked to approximate these values, rounding up to one year if their experience was less than one year.

\begin{table}[htb]
\centering
\caption{Descriptive Statistics of the Sample}
\label{tab:profpop}
\resizebox{\columnwidth}{!}{%
\begin{tabular}{cc|c|c|c|c}
\hline
\multicolumn{2}{l|}{}                 &  \textbf{Professional Experience (Years)} &  \textbf{Role Experience (Years)} & \textbf{Months in Project} &  \textbf{Age} \\ \hline
\multicolumn{2}{c|}{Mean}             & 7.81                                      & 5.19                              & 10.6                        & 29.62         \\ \hline
\multicolumn{2}{c|}{Median}           & 6                                         & 4                                 & 6                           & 28            \\ \hline
\multicolumn{2}{c|}{Minimum}          & 1                                         & 1                                 & 1                           & 18            \\ \hline
\multicolumn{2}{c|}{Maximum}          & 41                                        & 28                                & 126                         & 59            \\ \hline
\multicolumn{1}{c|}{\textbf{Percentiles}} & 25th & 3 & 2 & 3 & 25 \\ \cline{2-6} 
\multicolumn{1}{c|}{}                  & 50th & 6 & 4 & 6 & 28 \\ \cline{2-6} 
\multicolumn{1}{c|}{}                  & 75th & 10 & 7 & 12 & 33 \\ \hline
\end{tabular}
}
\end{table}

The sample consists of professionals with an average of approximately **eight years of experience**, of which **five years** were in their current role. On average, respondents had **11 months** in their current project and were **around 30 years old**.

A noteworthy observation is that **75\% of respondents had been in their current project for up to one year**, yet they had significant experience in their field. This pattern is further emphasized by the fact that **50\% had six months or less in their current project**, suggesting a **high turnover rate in software projects**.

Table~\ref{tab:descFuncao} presents the distribution of gender, job roles, and team types among the respondents.

\begin{table}[htb]
\centering
\caption{Distribution of Gender, Role, and Team Type in the Sample}
\label{tab:descFuncao}
\resizebox{\columnwidth}{!}{%
\begin{tabular}{lccc}
\hline
\multicolumn{2}{l|}{}                 & \multicolumn{1}{c|}{Frequency} & Percentage \\ \hline
 & \multicolumn{3}{c}{\textbf{Gender}}                           \\ \hline
 & \multicolumn{1}{c|}{Female}      & \multicolumn{1}{c|}{69}         & 16.7\%      \\ \hline
 & \multicolumn{1}{c|}{Male}     & \multicolumn{1}{c|}{344}        & 83.3\%      \\ \hline
 & \multicolumn{3}{c}{\textbf{Job Role}}                         \\ \hline
 & \multicolumn{1}{c|}{Analyst}      & \multicolumn{1}{c|}{71}         & 17.2\%      \\ \hline
 & \multicolumn{1}{c|}{Developer} & \multicolumn{1}{c|}{238}        & 57.6\%      \\ \hline
 & \multicolumn{1}{c|}{Tester}      & \multicolumn{1}{c|}{32}         & 7.7\%       \\ \hline
 & \multicolumn{1}{c|}{Manager}       & \multicolumn{1}{c|}{46}         & 11.1\%      \\ \hline
 & \multicolumn{1}{c|}{Other}        & \multicolumn{1}{c|}{25}         & 6\%       \\ \hline
 & \multicolumn{3}{c}{\textbf{Team Type}}                 \\ \hline
 & \multicolumn{1}{c|}{Agile}          & \multicolumn{1}{c|}{266}        & 64.4\%      \\ \hline
 & \multicolumn{1}{c|}{Hybrid}       & \multicolumn{1}{c|}{114}        & 27.6\%      \\ \hline
 & \multicolumn{1}{c|}{Traditional}   & \multicolumn{1}{c|}{33}         & 7.9\%       \\ \hline
\end{tabular}%
}
\end{table}

\subsubsection{Burnout Data}

Table~\ref{tab:descBurn} presents the **mean and standard deviation** of each burnout dimension. Respondents rated each burnout item using a **seven-point Likert scale**.

\begin{table}[htb]
\centering
\caption{Descriptive Statistics of Burnout Dimensions}
\label{tab:descBurn}
\resizebox{\columnwidth}{!}{%
\begin{tabular}{lrr}
\hline
\textbf{Burnout Dimension}   & \textbf{Mean} & \textbf{Standard Deviation} \\ \hline
\textbf{Emotional Exhaustion}  & 3.20  & 1.495  \\ 
\textbf{Cynicism}  & 2.43  & 1.39  \\ 
\textbf{Professional Efficacy (Inverse Scale)} & 5.58  & 1.010  \\ \hline
\end{tabular}%
}
\end{table}

Key observations:
- The **lowest-rated exhaustion item** was *"Working all day is really stressful for me"* (Mean = 1.67), while the **highest-rated exhaustion item** was *"I feel exhausted at the end of a workday"* (Mean = 4.03).
- **Cynicism was rated lower overall**, with the highest score in *"I feel like I am losing enthusiasm for my work"* (Mean = 3.43).
- **Professional efficacy was rated high**, indicating that respondents felt capable in their roles, which is inversely related to burnout.

These findings align with Cook \cite{cook2015job}, who reported that Software Engineers tend to experience **low levels of reduced self-efficacy**. However, unlike Cook’s study, our findings suggest that **emotional exhaustion is more concerning than cynicism in Software Engineering**.

\subsubsection{Instability Data}

Table~\ref{tab:descInsta} presents descriptive statistics for **perceived instability** in different dimensions.

\begin{table}[htb]
\centering
\caption{Descriptive Statistics of Instability Dimensions}
\label{tab:descInsta}
\resizebox{\columnwidth}{!}{%
\begin{tabular}{lrr}
\hline
\textbf{Instability Dimension}  & \textbf{Mean} & \textbf{Standard Deviation} \\ \hline
\textbf{Task Instability}  & 4.85  & 1.56  \\ 
\textbf{Team Instability}  & 3.92  & 1.65  \\ 
\textbf{Technological Instability} & 3.76  & 1.50  \\ \hline
\end{tabular}%
}
\end{table}

\section{Implications}
\label{sec:implications}

\subsection{Implications for Industry}

An interesting observation is the weak or non-existent correlations between instability dimensions and the **professional efficacy** dimension of burnout in software development. The data suggest that confidence in one's skills and work quality is **not strongly related to the perception of environmental changes**. However, **cynicism exhibits the highest levels of correlation with all three instability dimensions**.

A notable finding is that **perceptions of task, technological, and team instability are linked to feelings of exhaustion among software developers**. Based on this, the first managerial recommendation is:

\begin{quote}
    \textbf{Managers should implement strategies to reduce perceived instability among their teams, as higher instability correlates with increased burnout levels.}
\end{quote}

Additionally, since changes in the software development process are often **inevitable and frequent**, the second recommendation is:

\begin{quote}
    \textbf{Managers should actively monitor levels of exhaustion and cynicism to implement interventions that mitigate burnout.}
\end{quote}

Cynicism tends to increase in environments perceived as unstable. Cynicism, in turn, reflects **indifference and detachment from work**, as well as **reduced emotional engagement with colleagues and clients**. Thus, the third recommendation is:

\begin{quote}
    \textbf{In environments where instability increases, cynicism levels tend to rise, which can negatively impact team dynamics and work commitment. Strategies should be implemented to prevent this scenario.}
\end{quote}

One of the most concerning findings in this study is that the statement *"I feel exhausted at the end of the workday"* had the highest rating among burnout items. Therefore, the fourth recommendation is:

\begin{quote}
    \textbf{Organizations should implement initiatives to reduce end-of-day exhaustion and improve employee well-being.}
\end{quote}

While this may not be a definitive solution, it could help mitigate burnout and foster a healthier work environment.

\subsection{Implications for Academia}

From an academic perspective, these results provide **new insights into burnout in software development**. One key finding is that **reduced professional efficacy is less prevalent than the other burnout dimensions in Software Engineering**. Given that burnout dimensions are interrelated and can be leveraged to mitigate burnout as a whole \cite{maslach2016burnout}, the following research question emerges:

\begin{quote}
    \textit{How can high efficacy levels be leveraged to reduce the impact of other burnout dimensions in Software Engineering?}
\end{quote}

Another important consideration is whether the **general burnout scale** requires adaptations for the **Software Engineering field**. Other disciplines, such as healthcare and education, have developed specialized burnout instruments \cite{maslach2016burnout}. Although this study does not provide definitive evidence, further investigation is warranted to determine whether specific items should be revised for greater accuracy in software development contexts.

Additionally, given the **constant changes in software development**, future studies could explore how **instability** correlates with **job satisfaction, turnover intentions, and team conflict**.

Finally, **improving the measurement of technological instability** is a key research opportunity. Further studies should refine the **instability scale** and enhance its dimensions, given its relevance to the field.

\section{Conclusion}
\label{sec:conclusion}

The primary objective of this study was to **analyze the relationship between perceived instability and burnout in Software Engineering**. Specifically, the study sought to determine whether these relationships were **positive or negative**.

The findings indicate that **team, technological, and task instability have positive and significant relationships with exhaustion and cynicism**. However, for professional efficacy, the relationships were **negative and weak** for technological and team instability, and **non-existent** for task instability.

Regarding burnout, one of the most intriguing findings is the **high efficacy levels among software professionals**. This result, in conjunction with Cook’s findings \cite{cook2015job}, raises the following question:

\begin{quote}
    \textit{What factors contribute to the strong sense of professional efficacy among software engineers?}
\end{quote}

Finally, another consideration is whether certain **burnout scale items** require refinement for the **Software Engineering domain**. For instance, the item *"I just want to do my job without being disturbed"* may have different implications in software development, given the prevalence of **interruptions and dissatisfaction with them** \cite{latoza2006maintaining}.

\subsection{Limitations and Threats to Validity}

This section discusses the limitations encountered during this study. A key limitation of **survey-based research** is sample size, which affects **representativeness**. To mitigate this, the study leveraged social media networks of **influential figures** in the computing field to **maximize participation** and promote diverse responses across Brazil.

Another limitation is **inherent to the research design**. This study employed a **cross-sectional survey**, meaning that variables were measured at a **single point in time**. Consequently, responses **reflect the perceptions of respondents at the moment of data collection**, which may not capture long-term trends.

Additionally, **self-report measures** were used, which introduces potential biases. Some studies \cite{demetriou2015self} indicate that respondents may **overestimate their abilities** when self-assessing.

To address potential **validity concerns**, the study utilized a **widely validated burnout scale** in multiple disciplines, including Software Engineering. Moreover, reliability tests were conducted to verify scale consistency.

A key threat to validity lies in the **instability questionnaire**, which is still in its early validation phase and requires further refinement. Specifically, results related to **technological instability** should be interpreted with caution, as some factor loadings were **relatively low** (slightly above 0.400), and the **Cronbach's alpha for this dimension was the weakest**.

\section*{Artifact Availability}

The measurement scales used in this study have been **made available throughout the text**. However, due to confidentiality agreements made during data collection, the dataset is not publicly available. This policy will be revised in **future research projects**.

\bibliographystyle{ACM-Reference-Format}

\bibliography{referencias.bib}
\end{document}